\pdfoutput=1 
\documentclass[12pt,titlepage]{extarticle}
\usepackage{hyperref}
\usepackage[utf8]{inputenc}
\usepackage{cite}
\usepackage{float}
\usepackage{graphicx}
\usepackage{tikz}
\usepackage[title]{appendix}
\usepackage{epstopdf}
\usepackage{amsmath,amsfonts,amsthm,bm}
\usepackage{layout}
\usepackage{setspace}
\usepackage[font=scriptsize]{caption}
\usepackage{mathrsfs}
\usepackage[T1]{fontenc}
\usepackage[utf8]{inputenc}
\usepackage{authblk}
\usepackage[natbibapa]{apacite}
\numberwithin{equation}{section}
\title{\bf Financial Application of Extended Residual Coherence}
\author{Xuze Zhang}
\author{Benjamin Kedem}
\affil{Department of Mathematics and Institute for Systems
Research, University of Maryland, College Park}
\date{}
\begin{document}
\maketitle
\begin{abstract}

Residual coherence is a graphical tool for selecting potential second-order interaction terms as functions of a single time series and its lags. This paper extends the notion of residual coherence to account for interaction terms of multiple time series. Moreover, an alternative criterion, integrated spectrum, is proposed to facilitate this graphical selection. 

A financial market application shows that new insights can be gained regarding implied 
market volatility.
\end{abstract}

\section{Introduction}

Nonlinear phenomena in random processes have attracted much attention going back to the work of \cite{wiener} 
concerning random nonlinear oscillators excited by a random input,
random shot effect as input for testing nonlinear circuits, and more generally concerning a class of nonlinear polynomial
functionals to model input-output relationships in nonlinear systems. In Wiener's words he  was
interested in ``methods of handling the spectrum," which motivates the use of higher order spectra dealt with by quite a few authors including
\cite{brillinger1965}, \cite{BR1967}, \cite{hinich1979}, \cite{nikias1993}, and \cite{elgar1998}.
The excellent review paper by \cite{sanaullah2013} provides numerous additional references about applications of nonlinear
techniques based on higher order spectra.
Inherent in all nonlinear systems is the problem of assessing the degree and extent
of nonlinearity, which can be approached by the detection of nonlinear components or {\em interactions}
(\cite{tick1961}, \cite{elgar1998}).

In this paper the detection of nonlinear second-order interactions is done by an extension of {\em residual coherence}
introduced in \cite{K14} and applied in mortality forecasting. Residual coherence is
a nonlinear variation of the well known measure of linear coherence.
The method is then applied to two volatility indices,
the Chicago Board Options Exchange Volatility Index (VIX),
and the Russell 2000 Volatility Index (RVX).

\section{Extensions of residual coherence}

\subsection{Preliminaries}

The coherence between two time series $(X(t),Y(t))$ measures the extent of linear relationship between them in the
frequncy domain. Provided all auto- and cross-spectra exist, it is defined as
\begin{equation}
\gamma_{XY}(\lambda)=\frac{|f_{XY}(\lambda)|^2}{f_{XX}(\lambda)f_{YY}(\lambda)}
\end{equation}
(see \cite{koopmans}) where $f_{XX}$ and  $f_{YY}$ are the spectra of $X(t)$ and $Y(t)$, respectively, and $f_{XY}$
is the cross-spectral density of $X(t)$ and $Y(t)$.
This is widely used in
detecting connections and clustering of time series.
Relevant works include \cite{sun2004}, \cite{maharaj2010} and \cite{euan2019}, among many others. When the relationship is nonlinear, it is frequently analyzed by  bispectra, trispectra, or higher-order spectra. For example, a bispectral method for detecting
 lag processes was proposed by \cite{hinich1979}.
Lagged coherence and residual coherence were first introduced in \cite{K75} and \cite{K14}, respectively,
to detect and select potential interaction effects as input to nonlinear systems, based on an orthogonal decomposition in \cite{K74} without involving bispectrum or higher-order spectra.

Let $Y(t)$ be the output of a system of which the input consists of linear and quadratic filters of $X(t)$ plus noise $\varepsilon(t)$,
\begin{equation}
Y(t)=L[X(t)]+\sum_{k=1}^{\infty}L_{u_k}[\tilde{X}_{u_k}(t)]+\varepsilon(t)
\end{equation}
where $\tilde{X}_{u_k}$ is a lag process defined as $\tilde{X}_{u_k}(t)=X(t)X(t-u_k)-E[X(t)X(t-u_k)]$. For simplicity, assume that $Y(t)$ and $X(t)$ are zero-mean real valued processes and that
all relevant auto- and cross-spectra exist. Then, for sufficiently large $n$, $Y(t)$ can be approximated by
\begin{equation}
Y^{\ast}(t)=G_{1}(t)+\sum_{k=1}^{n}G_{2,k}(t)+\varepsilon(t)
\label{approx}
\end{equation}
where $G_{1}(t)$ is a linear filter of $X(t)$, and as in \cite{K75}, $G_{2,k}(t)$ is a sum of a linear filter of $X(t)$ and a linear filter of $\tilde{X}_{u_k}(t)$, such that $G_{2,k}(t)\bot G_{1}(t)$ for $k=1,\dots,n$.

\cite{K75} showed that if there is prior knowledge that $Y(t)$ takes on a simpler form
\begin{equation}
\label{1lag}
Y(t)=L[X(t)]+L_{u}[\tilde{X}_{u}(t)]+\varepsilon(t),
\end{equation}
then it can be rewritten as a sum of two orthogonal processes, $G_{1}(t)$ and $G_{2}(t;u)$ plus noise $\varepsilon(t)$,
\begin{equation}
Y(t)=G_{1}(t)+G_{2}(t;u)+\varepsilon(t).
\label{dec}
\end{equation}
Then the lag process, or interaction,
$\tilde{X}_{u}(t)$ that minimizes $E\varepsilon^{2}(t)$ can be selected by finding the
lag $u$ that maximizes the lagged coherence $S_{2}(\lambda;u)$ over all frequencies $\lambda\in[-\pi,\pi]$ such that
\begin{equation}
S_{2}(\lambda;u)=\frac{f_{G_{1}G_{1}}(\lambda)+f_{G_{2}G_{2}}(\lambda;u)}{f_{YY}(\lambda)}.
\end{equation}

However, as mentioned in \cite{K75}, there might not exists a $u$ that maximizes $S_{2}(\lambda;u)$ over all frequencies. One way to resolve this issue is to define the residual coherence as
\begin{equation}
RC(u)=\sup_{\lambda}\frac{f_{G_{2}G_{2}}(\lambda;u)}{f_{YY}(\lambda)}
\end{equation}
and find $u$ that maximizes $RC(u)$. This is shown to be useful for interaction selection in \cite{K14} and \cite{K16}.

\subsection{Lagged coherence and residual coherence for more than two orthogonal components}

Consider the model
\begin{equation}
\label{mod}
Y(t)=\sum_{k=1}^{n}L_{k,u_k}[X_{k,u_k}(t)]+\varepsilon(t).
\end{equation}
The goal is to select $X_{k,u_k}(t)$ from a certain family of processes $\{X_{k,u_k}(t):u_k=1, 2, \dots\}$ for $k=1,\dots,n$. 
This reduces to (\ref{1lag}) when $n=2$ and $L_{1,u_1}[X_{1,u_1}(t)]$ is the linear filter of $X(t)$.
For $n>2$, we shall extend the orthogonal decomposition (\ref{dec}) 
\begin{equation}
Y(t)=\sum_{k=1}^{n}G_{k}(t;u_{1},\dots,u_{k})+\varepsilon(t)
\end{equation}
where all $G_{k}$'s for $k=1,\dots,n$ are mutually orthogonal, to account for more orthogonal components, given by
\begin{equation}
\begin{split}
G_{k}(t;u_{1},\dots,u_{k})&=\sum_{j=1}^{k}\int_{-\pi}^{\pi}e^{it\lambda}A_{j,k-j+1}(\lambda)dZ_{X_{j,u_j}}(\lambda)
\end{split}
\end{equation}
for $k=1,\dots,n$, where the $A$'s are non-zero and $Z$'s are the corresponding spectral measures.

The $A$'s can be obtained by using the orthogonal conditions among $G_k$'s such that
\begin{equation}
\begin{split}
A_{k,1}(\lambda)&=\overline{\left[\frac{\sum_{j=1}^{k}c_{k,j}(\lambda)f_{X_{j,u_j}Y}(\lambda)}{\sum_{j=1}^{k}c_{k,j}(\lambda)f_{X_{j,u_j}X_{k,u_k}}(\lambda)}\right]}\phantom{aaa}k=1,\dots,n \\
\end{split}
\end{equation}
and
$$
A_{j,k-j+1}(\lambda)=c_{k,j}(\lambda)A_{k,1}(\lambda)\phantom{aaa}j=1,\dots,k, \phantom{aaa}k=1,\dots,n\\
$$
where 
\begin{equation*}
\begin{split}
c_{k,k}(\lambda)=1,\phantom{aa}c_{k,j}(\lambda)=\frac{\bm{F}_{k,j}(\lambda)}{\bm{F}_{k}(\lambda)},\phantom{aa}\bm{F}_{k}(\lambda)=(f_{i,j}(\lambda))_{(k-1)\times (k-1)},\phantom{aa}f_{i,j}\equiv f_{X_{j,u_j}X_{i,u_i}},
\end{split}
\end{equation*}
and $\bm{F}_{k,j}(\lambda)$ is equivalent to $\bm{F}_{k}(\lambda)$ of which $j$th column is replaced by 
\begin{equation*}
\bm{f}_{k}(\lambda)\equiv-[f_{1,k}(\lambda),\dots,f_{k-1,k}(\lambda)]^T. 
\end{equation*}
More details are provided in Appendix \ref{apd}.

Subsequently,
\begin{equation}
\label{est}
\begin{split}
f_{G_{k}G_k}(\lambda;u_1,\dots,u_k)&=|A_{k,1}(\lambda)|^{2}\left[\sum_{j=1}^{k}c_{k,j}(\lambda)f_{X_{j,u_j}X_{k,u_k}}(\lambda)\right]\\
S_{k}(\lambda;u_1,\dots,u_k)&=\frac{\sum_{j=1}^{k}f_{G_{j}G_j}(\lambda;u_1,\dots,u_j)}{f_{YY}(\lambda)}\\
RC(u_1,\dots,u_k)&=\sup_{\lambda}[S_{k}(\lambda;u_1,\dots,u_k)-S_{m}(\lambda;u_1,\dots,u_m)]\phantom{aa}1\leq m\leq k
\end{split}
\end{equation}
for $k=1,\dots,n$. Note that $S_{k}(\lambda;u_1,\dots,u_k)$ and $RS(u_1,\dots,u_k)$ depends only on $u_{m+1},\dots,u_k$ once $u_1,\dots,u_m$ are determined. The estimates of the above quantities are obtained based on the estimates of the relevant auto- and cross-spectra.

\subsection{Selection Criteria}

In this section, lagged coherence and residual coherence are examined and an alternative criterion is proposed. Take $n=2$ and fix $u_1$, then it reduces to the case in \cite{K75}. It illustrates that if there exists a $u_2$ that maximizes $S_{2}(\lambda;u_2)$ for all $\lambda$, then such $u_2$ minimizes $E\varepsilon^{2}(t)$ in
\begin{equation}
E\varepsilon^{2}(t)=\int_{-\pi}^{\pi}f_{\varepsilon\varepsilon}(\lambda)d\lambda=\int_{-\pi}^{\pi}f_{YY}(\lambda)[1-S_{2}(\lambda;u_2)]d\lambda.
\end{equation}
Indeed, the quantity we wish to maximize is $\int_{-\pi}^{\pi}f_{G_{2}G_{2}}(\lambda;u_2)d\lambda$ since
\begin{equation}
\int_{-\pi}^{\pi}f_{YY}(\lambda)S_{2}(\lambda;u_2)d\lambda=\int_{-\pi}^{\pi}[f_{G_{1}G_{1}}(\lambda)+f_{G_{2}G_{2}}(\lambda;u_2)]d\lambda
\end{equation}
based on (\ref{est}). Such criterion works even if such $u_2$ does not exist so that this can be an alternative to residual coherence. This criterion can be readily extended to a more general case. Suppose there is prior knowledge for the inclusion of first $m$ processes i.e. $u_1,\dots,u_m$ are fixed, then we define the integrated spectrum
\begin{equation}
IS(u_{m+1},\dots,u_n)\equiv\int_{-\pi}^{\pi}\sum_{k=m+1}^{n}f_{G_{k}G_{k}}(\lambda;u_{k},\dots,u_n)d\lambda
\end{equation}
and find $u_{m+1},\dots,u_n$ that maximizes $IS(u_{m+1},\dots,u_n)$.

Once all $u$'s are determined, the regression-based selection method proposed in \cite{K14} and \cite{K16} is used to select significant terms within the processes selected by the graphical method and this is illustrated in both Section \ref{sim} and Section \ref{apply}. Relevant regression for time series that the selection entails can be found in \cite{KF2002}.

\section{Simulation}
\label{sim}
In this section, a simulation is performed with $n=4$ and $u1,u2$ fixed to validate and compare the two criteria, residual coherence and integrated spectrum. The steps are as follows:\\
\begin{enumerate}
\item Generate $\{x_{1}(t)\}_{t=1}^{1010}$ from an AR(1) process $X_{1}(t)=0.4X_{1}(t-1)+u_{1}(t)$ and $\{x_{2}(t)\}_{t=1}^{1010}$ from an AR(1) process $X_{2}(t)=0.2X_{2}(t-1)+u_{2}(t)$ where $u$'s are white noise $N(0,1)$.
\item Obtain 
\begin{equation}
\label{yt}
y(t)=0.4x_{1}(t)+0.3x_{2}(t)+0.4x_{1}(t-2)x_{2}(t-1)+0.3x_{1}(t)x_{2}(t-4)+\varepsilon(t) 
\end{equation}
where $\varepsilon$'s are white noise $N(0,1)$ and t=11,\dots,1010 so that all relevant series have length 1000.

\item This is the model (\ref{mod}) with $n=4$ and known $X_{1}(t)$, $X_{2}(t)$. We considered selecting $X_{3,u_3}(t)$ and $X_{4,u_4}(t)$ from the family $\{X_{1}(t+h)X_{2}(t):h=-9,-8,\dots,0,\dots,9\}$. In fact, this family can be made larger and the choice here only serves as an example. Then we estimated all relevant auto- and cross-spectra using Tukey-Hamming kernel with window size 10 for frequencies $\lambda_{k}=-\pi+k\pi/1000$, $k=0,\dots,2000$. Subsequently, we estimated $RC(u_3)$, $IS(u_3)$, $RC(u_3,u_4)$ and $IS(u_3,u_4)$ for $u_3,u_4=-9,-8,\dots,0,\dots,9$. Note that $\widehat{IS}(u_3)=\sum_{k=1}^{2000}\pi\hat{f}_{G_{3}G_{3}}(\lambda_k;u_3)/1000$ and $\widehat{IS}(u_3,u_4)=\sum_{k=1}^{2000}\pi[\hat{f}_{G_{3}G_{3}}(\lambda_k;u_3)+\hat{f}_{G_{4}G_{4}}(\lambda_k;u_3,u_4)]/1000$.
\end{enumerate}

\begin{figure}[htbp]
\centering
\includegraphics[width=4.5in]{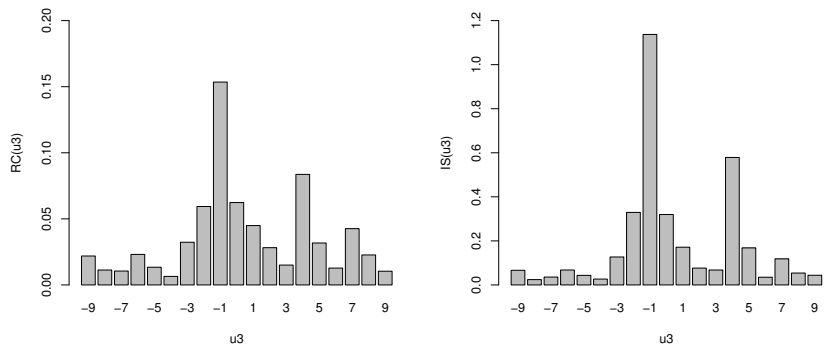}
\caption{$\widehat{RC}(u_3)$ (left) and $\widehat{IS}(u_3)$ (right) for $u_3=-9,\dots,9$.}
\label{Fig1}
\end{figure}

\begin{figure}[htbp]
\centering
\includegraphics[width=4.5in]{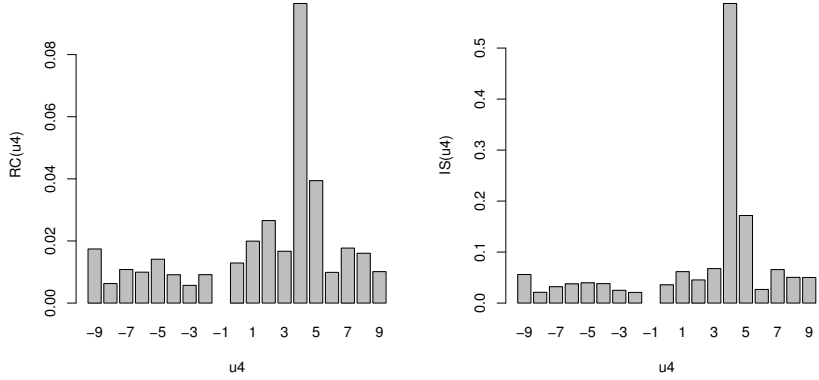}
\caption{$\widehat{RC}(u_4)$ (left) and $\widehat{IS}(u_4)$ (right) for $u_4=-9,\dots,9$. Note that the bars that correspond to $u_4=-1$ are set to be 0 since $u_3=-1$.}
\label{Fig2}
\end{figure}

The results are shown by Figure \ref{Fig1}
which indicates that the process $X_{1}(t-1)X_{2}(t)$ is the optimal choice for the third input. It is also observed from Figure \ref{Fig1} that $X_{1}(t+4)X_{2}(t)$ is another potential input since the bars that correspond to $u_3=4$ are the second highest ones in both graphs. With $u_3=-1$ fixed, $u_4$ can be determined by $\widehat{RC}(u_4)$ and $\widehat{IS}(u_4)$ as shown by Figure \ref{Fig2}
and both graphs indicate that $u_4=4$ is the optimal choice, which accords with the original model (\ref{yt}).

With $u_3$ and $u_4$ determined, we select significant covariates from the selected processes $X_{1}(t-1)X_{2}(t)$ and $X_{1}(t+4)X_{2}(t)$ using the regression-based method propsed in \cite{K14} and \cite{K16}. We selected four lag terms from each input i.e. $x_{1}(t), \dots, x_{1}(t-3),x_{2}(t),\dots,x_{2}(t-3),x_{1}(t-1)x_{2}(t),\dots,x_{1}(t-4)x_{2}(t-3),x_{1}(t)x_{2}(t-4),\dots,x_{1}(t-3)x_{2}(t-7)$ and regressed $y(t)$ on all the selected covariates. We then performed stepwise selection based on AIC and it is observed from Table \ref{regres} that the selected model is similar to (\ref{yt}) since the significant ($\alpha=0.05$) covariates are identical to the ones in (\ref{yt}) and the estimated coefficients correspond to the ones in (\ref{yt}), which validates the method.

\begin{table}[H]
\centering
\captionsetup{margin=10pt,font=small,labelfont=bf}
\caption{Regression result of the selected model}
\scalebox{0.9}{
\begin{tabular}{lccc}
\hline
  & Estimate & SE & p-value\\
 \hline
Intercept & -0.0264 & 0.0322 & 0.4125 \\
$x_{1}(t)$ &0.3876 &0.0292 & 0.0000\\
$x_{2}(t)$ & 0.2907 & 0.0303 & 0.0000 \\
$x_{2}(t-2)$ & 0.0571 & 0.0307& 0.0629 \\
$x_{2}(t-3)$ & -0.0555& 0.0307 & 0.0708\\
$x_{1}(t-2)x_{2}(t-1)$ & 0.3729 & 0.0277& 0.0000\\
$x_{1}(t)x_{2}(t-4)$& 0.2569 & 0.0275& 0.0000\\
 \hline
\end{tabular}
}
\label{regres}
\end{table}

Remark: It seems that regression-based selection criteria of interaction terms can be applied directly,
thus bypassing the need for our graphical method. However, we rationalize the use of our
spectral graphical
selection for the following reason.
The number of potential covariates in the initial
model might be too large which could result in conflicting selections
and possible inconsistencies depending on
the model selection method.
Our graphical method identifies potentially useful interactions which can then
be taken into account  and reduce significantly the number of covariates fed into any
model selection method, thus rendering the selection more manageable.

\section{An Application to Volatility Index}
\label{apply}

The Volatility Index of a certain underlying asset gives the expectation of the corresponding market volatility in a certain future period. The first and most famous one, the Chicago Board Options Exchange Volatility Index (VIX), was introduced by \cite{whaley1993}. The underlying asset for VIX is the S\&P 500 index so that it reflects the implied volatility of the stock performance of large capitalization companies. For the implied volatility of small capitalization stocks,
we have chosen the Russell 2000 Volatility Index (RVX). These two volatility indices shall be considered here
as indicators for the stock market. For commodity markets, two important volatility indices, the Crude Oil Exchange Traded Funds Volatility Index (OVX) and the Gold Exchange Traded Funds Volatility Index (GVX) were used. The two-year (2018-2019) daily data of these four series were taken from the Federal Reserve Economic Data Website (https://fred.stlouisfed.org/).

The above methods were applied to analyze the relationships between the volatility indices of the stock and commodity markets. This section is divided into two parts, one investigates the influence of OVX and GVX on VIX and the other examines the influence on RVX.

\begin{figure}[htbp]
\centering
\includegraphics[width=5in]{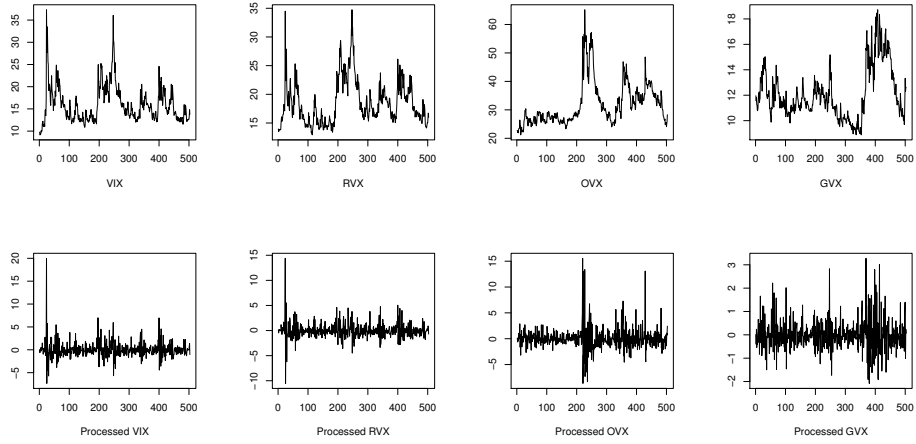}
\caption{VIX, RVX, OVX and GVX series before and after processing}
\label{Fig3}
\end{figure}

Before the analysis,
the four series were pre-processed to render them
approximately stationary. That was achieved by first-order differencing  
of the original series and centering  at 0.
Figure \ref{Fig3} depicts the four series before and after processing.

\subsection{VIX, OVX and GVX}
\label{VOG}
Consider the processed VIX series as the output and the processed OVX and GVX series as the input. Denote the processed VIX as $y(t)$, processed OVX as $x_{1}(t)$, and processed GVX as $x_{2}(t)$. The results of linear regression
of $Y(t)$ on $x_{1}(t)$ and $x_{2}(t)$ in Table \ref{regx} indicate that it is reasonable to include these two series as input since their coefficients are significant. Note that the intercept is omitted since all three series were centered at 0.

\begin{table}[H]
\centering
\captionsetup{margin=10pt,font=small,labelfont=bf}
\caption{Regression result of $y(t)$ on $x_{1}(t)$ and $x_{2}(t)$ }
\scalebox{0.9}{
\begin{tabular}{lccc}
\hline
  & Estimate & SE & p-value\\
 \hline
$x_{1}(t)$ &0.1486 &0.0347 & 0.0000\\
$x_{2}(t)$ & 0.9395& $0.1075$ & 0.0000 \\
 \hline
\end{tabular}
}
\label{regx}
\end{table}

\begin{figure}[htbp]
\centering
\includegraphics[width=4.5in]{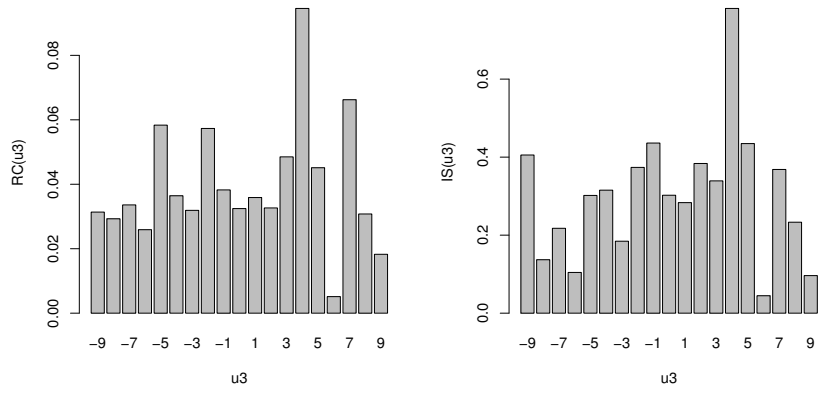}
\caption{$\widehat{RC}(u_3)$ (left) and $\widehat{IS}(u_3)$ (right) for $u_3=-9,\dots,9$.}
\label{Fig4}
\end{figure}

\begin{figure}[htbp]
\centering
\includegraphics[width=4.5in]{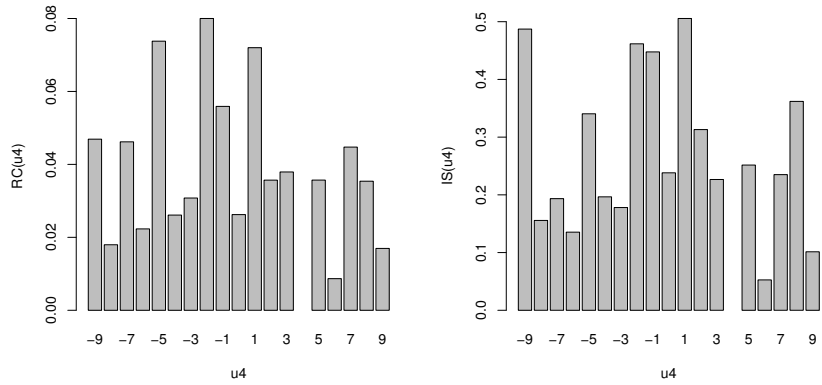}
\caption{$\widehat{RC}(u_4)$ (left) and $\widehat{IS}(u_4)$ (right) for $u_4=-9,\dots,9$. Note that the bars that correspond to $u_4=4$ are set to be 0 since $u_3=4$}
\label{Fig5}
\end{figure}

Then the goal is to find the third input based on the cross products of $x_{1}(t)$ and $x_{2}(t)$. This resembles the simulation problem so that we perform the same analysis as we did in Section \ref{sim}. We first select the third input from the family of processes $\{X_{1}(t+h)X_{2}(t):h=-9,-8,\dots,0,\dots,9\}$ and the estimated $RC$'s and $IS$'s are shown in Figure \ref{Fig4}
and it is observed that both criteria indicate that $u_3=4$ is the optimal choice. The $u_4$ is checked with $u_3=4$ fiexd and  \ref{Fig5} shows that none of the bars is particularly prominent so that we stop at the third input. 
In correspondence to Section \ref{sim}, we selected four lag terms from each of the three input series and performed a stepwise selection. The final model selected by AIC is shown in Table \ref{regres2}. It includes
the two significant ($\alpha$=0.05) interaction terms $x_{1}(t)x_{2}(t-4)$ and $x_{1}(t-1)x_{2}(t-5)$.

\begin{table}[H]
\centering
\captionsetup{margin=10pt,font=small,labelfont=bf}
\caption{Regression result of the selected model}
\scalebox{0.9}{
\begin{tabular}{lccc}
\hline
  & Estimate & SE & p-value\\
 \hline
Intercept & -0.0007 & 0.0717& 0.9927 \\
$x_{1}(t)$ &0.1677 &0.0360 & 0.0000\\
$x_{2}(t)$ & 0.9039 & 0.1133 & 0.0000 \\
$x_{2}(t-1)$ & -0.2504 & 0.1078& 0.0206 \\
$x_{1}(t)x_{2}(t-4)$  & 0.1344& 0.0498 & 0.0072\\
$x_{1}(t-1)x_{2}(t-5)$ & -0.1010 & 0.0508& 0.0473\\
$x_{1}(t-2)x_{2}(t-6)$& -0.0914 & 0.0504& 0.0707\\
$x_{1}(t-3)x_{2}(t-7)$& 0.0817 & 0.0501& 0.1037\\
 \hline
\end{tabular}
}
\label{regres2}
\end{table}

\subsection{RVX, OVX and GVX}
\begin{figure}[htbp]
\centering
\includegraphics[width=4.5in]{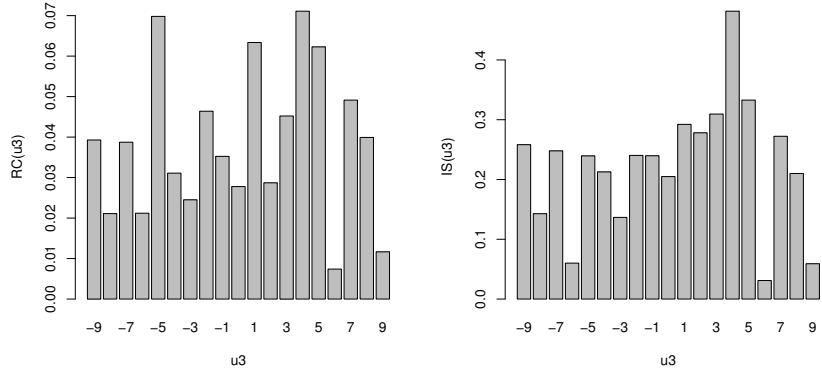}
\caption{$\widehat{RC}(u_3)$ (left) and $\widehat{IS}(u_3)$ (right) for $u_3=-9,\dots,9$.}
\label{Fig6}
\end{figure}

\begin{figure}[htbp]
\centering
\includegraphics[width=4.5in]{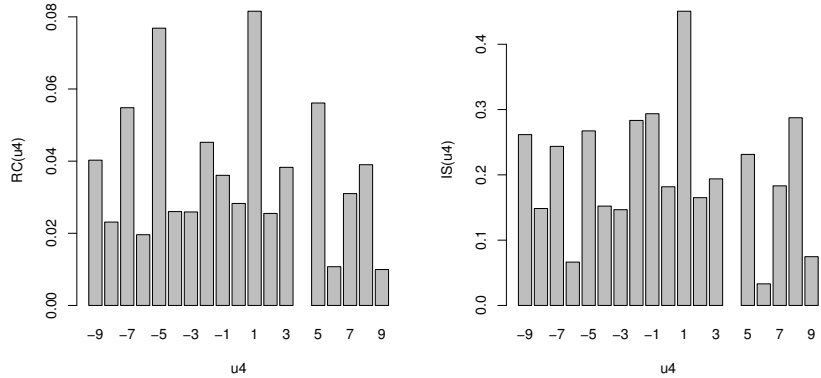}
\caption{$\widehat{RC}(u_4)$ (left) and $\widehat{IS}(u_4)$ (right) for $u_4=-9,\dots,9$. Note that the bars that correspond to $u_4=4$ are set to be 0 since $u_3=4$}
\label{Fig7}
\end{figure}

We repeated the analysis in Section \ref{VOG} with VIX replaced by RVX. We still consider the
processed OVX and GVX as input and try to detect possible significant
interactions. In Figure \ref{Fig6}, both bar plots indicate that the optimal choice for $u_3$ is 4 while the
bar plot of $\widehat{RC}(u_3)$ indicates that we might need to consider $1$ and $-5$ as well.
Therefore, we checked for $u_4$ and Figure \ref{Fig7} shows that no bar stands out in the graph of $\widehat{RC}(u_4)$ while the bar of $u_4=1$ is prominent in the graph $\widehat{IS}(u_4)$. Therefore, we took $X_{1}(t+1)X_{2}(t)$ as the fourth input. 
We selected the lag terms as in Section \ref{sim} and \ref{VOG} and the result of stepwise regression based on AIC is shown in Table \ref{regres3}. Four significant ($\alpha=0.05$) interation terms, $x_{1}(t)x_{2}(t-4)$, $x_{1}(t-1)x_{2}(t-5)$, $x_{1}(t)x_{2}(t-1)$ and $x_{1}(t-1)x_{2}(t-2)$, are detected where the first two are from
$X_{1}(t+4)X_{2}(t)$ and the last two are from $X_{1}(t+1)X_{2}(t)$.

\begin{table}[H]
\centering
\captionsetup{margin=10pt,font=small,labelfont=bf}
\caption{Regression result of the selected model}
\scalebox{0.9}{
\begin{tabular}{lccc}
\hline
  & Estimate & SE & p-value\\
 \hline
Intercept & 0.0035 & 0.0626& 0.9549 \\
$x_{1}(t)$ &0.1253 &0.0314 & 0.0001\\
$x_{2}(t)$ & 0.8296 & 0.0994 & 0.0000 \\
$x_{2}(t-1)$ & -0.2130 & 0.0954& 0.0261 \\
$x_{1}(t)x_{2}(t-4)$  & 0.1191& 0.0440 & 0.0071\\
$x_{1}(t-1)x_{2}(t-5)$ & -0.1114 & 0.0448& 0.0132\\
$x_{1}(t-3)x_{2}(t-7)$& 0.0632 & 0.0432& 0.1438\\
$x_{1}(t)x_{2}(t-1)$& 0.1049 & 0.0512& 0.0410\\
$x_{1}(t-1)x_{2}(t-2)$& -0.1313 & 0.0497& 0.0085\\
 \hline
\end{tabular}
}
\label{regres3}
\end{table}

\section{Conclusion}

Residual coherence and integrated spectrum proposed in this paper are graphical devices which point to possible significant interactions
based on the result of Sections \ref{sim} and  \ref{apply}. Significant interactions
could produce one or more than one prominent bars in the bar plots of $RC(u_k)$ and $IS(u_k)$ as functions of the $k$th input
interaction.

When there are multiple prominent bars, one could consider $u_{k+1}$ for more possible significant interactions.
Once the input processes are determined, one can employ the regression-based selection method proposed in \cite{K14} and
\cite{K16} to search for significant covariate interactions.

In addition, it is observed from the analysis in Section \ref{apply} that the cross product
interaction $X_{1}(t+4)X_{2}(t)$ of the first order differences of OVX and GVX has significant influence on the first order differences of VIX and RVX. This suggests that daily increments of implied volatility of the stock market is possibly influenced by products of the daily increments (and their lags) of implied volatility of commodity markets. The process $X_{1}(t+4)X_{2}(t)$ might be an essential factor in the relationship between the implied volatility of stock market and certain commodity markets and therefore further exploration is warranted .

\clearpage
\begin{appendices}
\section{}
\label{apd}
Since all $G_k$'s are mutually orthogonal, fix $k$, then $\forall h$,
\begin{equation}
\begin{split}
&EG_{k}(t+h;u_{1},\dots,u_{k})\overline{G_{j}(t;u_{1},\dots,u_{j})}=0\phantom{aaa}j=1,\dots,k-1\\
\Rightarrow&\int_{-\pi}^{\pi}e^{ih\lambda}\sum_{l=1}^{k}A_{l,k+l-1}(\lambda)f_{X_{l,u_l}X_{j,u_j}}(\lambda)d\lambda=0\phantom{aaa}j=1,\dots,k-1\\
\Rightarrow&\sum_{l=1}^{k}A_{l,k+l-1}(\lambda)f_{X_{l,u_l}X_{j,u_j}}(\lambda)d\lambda=0\phantom{aaa}j=1,\dots,k-1\\
\Rightarrow&\bm{F}_{k}(\lambda)\bm{A}_{k}(\lambda)=\bm{f}_{k}(\lambda)A_{k,1}(\lambda)
\end{split}
\end{equation}
where $\bm{A}_{k}(\lambda)\equiv[A_{1,k}(\lambda),A_{2,k-1},\dots,A_{k-1,2}]^T$. Then by Cramer's rule,
\begin{equation}
A_{j,k-j+1}(\lambda)=\frac{\bm{F}_{k,j}(\lambda)}{\bm{F}_{k}(\lambda)}A_{k,1}(\lambda)=c_{k,j}(\lambda)A_{k,1}(\lambda)
\end{equation}
for $j=1,\dots,k-1$.
Based on the orthogonality and the uniqueness of Fourier transform, we also have
\begin{equation}
\begin{split}
EG_{k}(t+h;u_{1},\dots,u_{k})\overline{Y(t)}&=EG_{k}(t+h;u_{1},\dots,u_{k})\overline{G_{k}(t;u_{1},\dots,u_{k})}\\
\sum_{j=1}^{k}A_{k,k-j+1}(\lambda)f_{X_{j,u_j}Y}(\lambda)&=\overline{A_{k,1}(\lambda)}\sum_{j=1}^{k}A_{k,k-j+1}(\lambda)f_{X_{j,u_j}X_{k,u_k}}(\lambda)\\
A_{k,1}(\lambda)\sum_{j=1}^{k}c_{k,j}(\lambda)f_{X_{j,u_j}Y}(\lambda)&=|A_{k,1}(\lambda)|^{2}\sum_{j=1}^{k}c_{k,j}(\lambda)f_{X_{j,u_j}X_{k,u_k}}(\lambda)\\
A_{k,1}(\lambda)&=\overline{\left[\frac{\sum_{j=1}^{k}c_{k,j}(\lambda)f_{X_{j,u_j}Y}(\lambda)}{\sum_{j=1}^{k}c_{k,j}(\lambda)f_{X_{j,u_j}X_{k,u_k}}(\lambda)}\right]}
\end{split}
\end{equation}
Therefore, all $A$'s for $G_k$ are solved. 
\end{appendices}
\clearpage
\bibliography{RCMS}
\end{document}